\documentclass[journal]{IEEEtran}
\usepackage{cite}
\usepackage{graphicx}
\usepackage{footnote}
\usepackage{bbding}
\usepackage{pifont}
\usepackage{wasysym}
\usepackage{amssymb}
\usepackage{multirow}
\usepackage[table,xcdraw]{xcolor}
\usepackage{diagbox}
\usepackage[flushleft]{threeparttable}
\makesavenoteenv{tabular}
\usepackage{tabu}
\usepackage{longtable}
\usepackage{cite}
\usepackage{epstopdf}
\usepackage{color}
\usepackage{enumitem}
\usepackage{booktabs}
\usepackage{subfigure}

\setlist{parsep=0pt,listparindent=\parindent}

\ifCLASSINFOpdf
\else
\fi

\begin{document}
%
\title{N2VSCDNNR: A Local Recommender System Based on Node2vec and Rich Information Network}

\author{Jinyin Chen, Yangyang Wu, Lu Fan, Xiang Lin, Haibin Zheng, Shanqing Yu, Qi Xuan,~\IEEEmembership{Member,~IEEE}
\thanks{This work is partially supported by Zhejiang Natural Science Foundation(LY19F020025), Integration and Application of Human-computer Fusion Intelligent Image Recognition Algorithm (2018B10063), National Natural Science Foundation of China (61502423, 61572439), Engineering Research Center of Cognitive Healthcare of Zhejiang Province, Zhejiang Science and Technology Plan Project (LGF18F030009, 2017C33149), and Zhejiang Outstanding Youth Fund (LR19F030001), Key technologies, system and application of Cyberspace Big Search, Major project of Zhejiang Lab (2019DH0ZX01). \emph{(Corresponding author: Qi Xuan.)}
}
\thanks{J. Chen, Y. Wu, X. Lin, H. Zheng, S. Yu and Q. Xuan are with the Institute of Cyberspace Security, and the College of Information Engineering, Zhejiang University of Technology, Hangzhou 310023, China (e-mail: \{chenjinyin, 2111603080, 201403080215, 201303080231, yushanqing, xuanqi\}@zjut.edu.cn).}
\thanks{J. Chen, S. Yu and Q. Xuan are also with the Zhejiang Lab, Hangzhou 311121, China.}
\thanks{L. Fan is with the Hong Kong Polytechnic University, 11 Yuk Choi Road, Hung Hom, Kowloon, Hong Kong SAR (e-mail: lufan@polyu.edu.hk).}
}


\maketitle

\begin{abstract}
Recommender systems are becoming more and more important in our daily lives. However, traditional recommendation methods are challenged by data sparsity and efficiency, as the numbers of users, items, and interactions between the two in many real-world applications increase fast. In this work, we propose a novel clustering recommender system based on node2vec technology and rich information network, namely N2VSCDNNR, to solve these challenges. In particular, we use a bipartite network to construct the user-item network, and represent the interactions among users (or items) by the corresponding one-mode projection network. In order to alleviate the data sparsity problem, we enrich the network structure according to user and item categories, and construct the one-mode projection category network. Then, considering the data sparsity problem in the network, we employ node2vec to capture the complex latent relationships among users (or items) from the corresponding one-mode projection category network. Moreover, considering the dependency on parameter settings and information loss problem in clustering methods, we use a novel spectral clustering method, which is based on dynamic nearest-neighbors (DNN) and a novel automatically determining cluster number (ADCN) method that determines the cluster centers based on the normal distribution method, to cluster the users and items separately. After clustering, we propose the two-phase personalized recommendation to realize the personalized recommendation of items for each user. A series of experiments validate the outstanding performance of our N2VSCDNNR over several advanced embedding and side information based recommendation algorithms. Meanwhile, N2VSCDNNR seems to have lower time complexity than the baseline methods in online recommendations, indicating its potential to be widely applied in large-scale systems.
\end{abstract}


\IEEEpeerreviewmaketitle

\section{Introduction}
\IEEEPARstart{D}{ue} to the fast development of E-commerce, nowadays more and more items are sold online. Although convenient, it is becoming more time-consuming as the diversity of items increases. Recommender systems~\cite{yang2017collaborative,lo2018temporal} thus have been developed to help people find the items they are interested in and save their time in the searching process. Recommender systems could efficiently avoid information overload, a problem caused by the increasing amount of data overwhelmingly. It can efficiently predict~\cite{fu2018link,Chen2018Fast,chen2018link} the likely preferences of the users, recommend related items for them to facilitate further decision.

One of the most critical issues for recommender systems is the data sparsity. With the increasing scale of the system, the number of items often reaches millions, even billion, leading to the quite less possibility of two users focusing on the same items. The common strategy to alleviate the data sparsity problem is using clustering-based recommendation~\cite{west2016recommendation}, also known as local recommendation. Clustering-based recommendation approaches tackle the sparsity challenge by compressing the sparse network into a series of subsets. They are much more general and could be easily implemented across domains. The first clustering-based recommendation method was proposed by Ungar et al.~\cite{ungar1998clustering}, where the authors proposed a statistical clustering model and determine suitable parameters based on  comparing different methods. In recent years, clustering-based recommendation methods~\cite{xue2005scalable,yu2014personalized} attracts lots of attention from researchers. 
Typically, there are two kinds of clustering-based methods, i.e., single-set clustering and co-clustering. Single-set clustering methods, such like user clustering~\cite{esslimani2009collaborative,Joseph2012Beyond,rana2014evolutionary} and item clustering~\cite{o1999clustering}, cluster variables separately; while co-clustering recommenders~\cite{george2005scalable,zhang2013localized,xu2012exploration} cluster users and items simultaneously. By comparison, single-set clustering methods are more feasible to exploit side-information, while co-clustering methods focus on transaction information~\cite{deodhar2007framework}. 
In this work, we thus propose a single-set clustering-based recommendation framework to integrate both network topology and side-information.

Recently, network analysis technologies are becoming more and more popular for complex systems~\cite{xuan2018modern}. Such technologies can alleviate the data sparsity problem~\cite{xuan2015temporal} to certain extent and capture latent information beyond explicit features. In recommender systems, it is natural to represent user-item relationships as bipartite networks. However, there are relatively few network analysis methods are designed to directly analyze bipartite networks~\cite{xuan2009empirical}. In this paper, similar to~\cite{zhou2007bipartite}, we first construct the corresponding one-mode projection category networks which only contains users and items, respectively. Then we apply a special network embedding method on these one-mode networks. In particular, we employed node2vec~\cite{grover2016node2vec}, an advanced network representational learning algorithm, to automatically extract low-dimensional vectors for nodes. This method can capture both local and global structural information. Moreover, to overcome information loss in the process of projection and representation, we integrate category information to build a more informative network.

In particular, we first construct two basic bipartite networks, i.e., user-item network and item-category network, based on which we further generate a user-category network. Then, we transform the three bipartite networks to two one-mode projection networks. Next, based on the two related one-mode projection networks, we use node2vec algorithm~\cite{grover2016node2vec} to map each node into a vector, and use our SCDNN~\cite{chen2017improved} to cluster users and items separately, aiming to deal with the information loss problem of the projection. Finally, we proposed the two-phase personalized recommendation to realize the personalized recommendation of items for each user. By comparing with~\cite{jinyin2017improved}, the main contributions of this paper are as follows.
\begin{itemize}
\item We first propose a novel network embedding based clustering recommender system by integrating item category as side information, namely N2VSCDNNR, which can alleviate the data sparsity problem to certain extent.
\item Then, we proposed a novel automatically determining cluster number method in SCDDN, which uses the normal distribution method to extract the information of the data points and determines the cluster centers based on the confidence interval principle.
\item Finally, our experimental results demonstrate the outstanding performance of N2VSCDNNR over several advanced embedding and side information based recommendation algorithms, and meanwhile N2VSCDNNR has relatively lower time complexity than the others, making it suitable for online recommendations.
\end{itemize}

The remainder of the paper is organized as follows. In Sec.~\ref{RW}, we present the related works on clustering-based and embedding-based recommender systems; In Sec.~\ref{SCDNN}, we propose our N2VSCDNNR, which is based on both network embedding and clustering algorithms while integrates item category as side information; In Sec.~\ref{Exp}, we compare our N2VSCDNNR with the previous version N2VSCDNN~\cite{chen2017improved} and several advanced network embedding or side information based recommendation algorithms on multiple real-world datasets; Finally, we conclude the paper in Sec.~\ref{Conclusion}.

\begin{figure*}[!t]
\includegraphics[width=\linewidth]{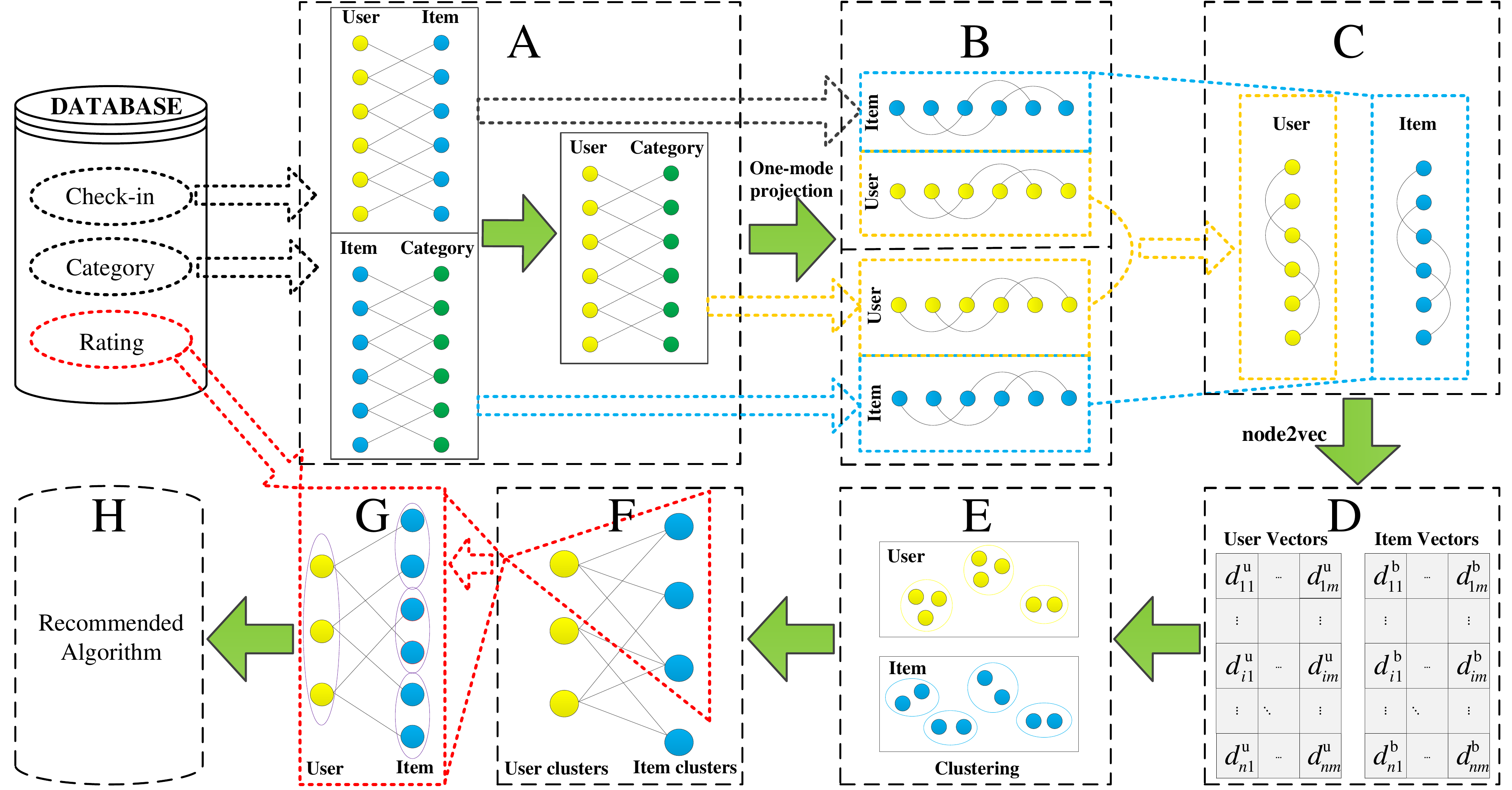}
\centering
\caption{The framework of N2VSCDNNR.}
\label{fig:frame}
\end{figure*}

\section{Related Works\label{RW}}
\subsection{Clustering-based Recommender Systems}
During the past two decades, a large number of studies on recommendation have emerged. Many recommendation methods suffer sparsity and scalability problems. Clustering-based recommendation methods thus have been widely developed to overcome such shortcomings to certain extent, where users (items) are grouped into multiple classes, providing a novel way to identify the nearest-neighbors.

Most of clustering-based recommendation methods compute the similarity based on rating data, and then employ some basic clustering algorithm, such as K-means method, to generate the users (items) groups. Sarwar et al.~\cite{sarwar2002recommender} proposed the bisecting K-means method clustering algorithm to divide the users into multiple clusters. In this method, the nearest-neighbors of a target user is selected based on the partition that the user belongs to. Puntheeranurak and  Tsuji~\cite{puntheeranurak2007multi} proposed a hybrid recommender system, where they clustered users by adopting a fuzzy K-means clustering algorithm. The recommendation results for both original and clustered data are combined to improve the traditional collborative filtering (CF) algorithms. Rana proposed a dynamic recommender system (DRS) to cluster users via an evolutionary algorithm. Wang clustered users by using K-means algorithm and then estimated the absent rating in the user-item matrix to predict the preference of a target user. Ji et al.~\cite{Ji2016Improving} paid more attention to discover the implicit similarity among users and items, where the authors first clustered user (or item) latent factor vectors into user (or item) cluster-level factor vectors. After that, they compressed the original approximation into a cluster-level rating-pattern based on the cluster-level factor vectors.

\subsection{Embedding-based Recommender Systems}
In network science, an important question is how to properly represent the network information. Network representation learning, used to learn low-dimensional representations for nodes or links in the network, is capable to benefit a wide range of real-world applications, such as recommender system~\cite{gao2018bine,palumbo2017entity2rec,dong2017metapath2vec,wen2018network,grad2017graph,wang2017music,palumbo2018knowledge}.

Recently, the DeepWalk algorithm~\cite{perozzi2014deepwalk} was proposed to transform each node in a network to a vector automatically, which takes full advantage of the information of the random walk sequence in the network. Another network representation learning algorithm based on simple neural network is the LINE algorithm~\cite{Tang2015LINE} which can be applied to large-scale directed weighted networks. Moreover, Grover and Leskovec~\cite{grover2016node2vec} suggested that increasing the flexibility in searching for adjacent nodes is the key to enhance network feature learning. They thus proposed the node2vec algorithm,  which learns low-dimensional representations for nodes by optimizing an objective of neighborhood preserving. It designs a flexible neighborhood sampling and a flexible biased random walk procedure that can explore neighborhoods through breadth-first sampling (BFS)~\cite{yang2014overlapping} or depth-first sampling (DFS)~\cite{henderson2012rolx}. It defines a second-order random walk with two parameters guiding the walk. One controls how fast the walk explores and the other controls how fast it leaves the neighborhood of the starting node. These two parameters allow our search to interpolate between BFS and DFS and thereby reflect an affinity for different notions of node equivalences.

Then, with the development of network embedding, a number of embedding-based recommender systems were proposed in recently years. For instance, Palumbo et al.~\cite{palumbo2017entity2rec} proposed an entity2rec algorithm to learning user-item relatedness from knowledge graphs, so as to realize item recommendation. Kiss and Filzmoser~\cite{grad2017graph} proposed a method to map the users and items to the same two-dimensional embedding space to make the recommendations. Swami~\cite{dong2017metapath2vec} introduced a heterogeneous representation learning model, called Metapath2vec++, which uses meta-path-based random walks to construct the heterogeneous neighbors of a node and then leverages a heterogeneous skip-gram model to perform node embeddings, and further make recommendations based on the network representation. Gao et al.~\cite{gao2018bine} proposed a network embedding method for bipartite networks, namely BiNE. It generates node sequences that can well preserve the long-tail distribution of nodes in the bipartite networks by performing biased random walks purposefully. The authors make recommendations with the generated network representation. Wen et al.~\cite{wen2018network} proposed an embedding based recommendation method. In this model, they use a network embedding method to map each user into a low dimensional space at first, and then incorporate user vectors into a matrix factorization model for recommendation.

\section{The Framework of N2VSCDNNR\label{SCDNN}}
In this paper, in order to recommend the items for the users more accurately, we propose N2VSCDNNR, with its whole framework shown in Fig.~\ref{fig:frame}, which consists of the following four steps.
\begin{enumerate}
\item Construct two bipartite networks, i.e., user-item network and item-category network, based on which we further generate a user-category network. Then, compress the three bipartite networks by one-mode projection to generate user-user projection network and item-item projection network, as shown in Fig.~\ref{fig:frame} A, B, and C.
\item Apply the node2vec algorithm to generate the user vectors and the item vectors according to the user-user projection network and item-item projection network, respectively, as shown in Fig.~\ref{fig:frame} D.
\item Use the SCDNN algorithm to cluster the user vectors and the item vectors into multiple clusters, respectively, as shown in Fig.~\ref{fig:frame} E.
\item Use the two-phase personalized recommendation to recommend suitable items to users. First, recommend item-clusters to each user-cluster based on K-means method, as shown in Fig.~\ref{fig:frame} F; Second, realize the personalized recommendation of items for each user in the user cluster, as shown in Fig.~\ref{fig:frame} G and H.
\end{enumerate}

\subsection{One-mode Projection of Bipartite Networks}
In recommender systems, constructing user-item bipartite networks based on their relationships is ubiquitous. But there is always data sparsity problem, i.e., part of users have very little records, making the constructed bipartite network not sufficient to capture the real relationship between users and items. We thus introduce item categories to effectively solve the sparsity problem. In particular, the one-mode projections of bipartite networks are performed by the following four steps.
\begin{enumerate}
\item Build two bipartite networks, i.e., the user-item bipartite network and the item-category bipartite network, as shown in Fig.~\ref{fig:frame} A.
\item Build the user-category bipartite network by integrating the user-item bipartite network and the item-category bipartite network, as shown in Fig.~\ref{fig:frame} A, where the weight between a user and a category is the total number of times that the user check the items in this category.
\item Project the user-item network into two separate networks, i.e., a user-user network and an item-item network, where the weight $CK_{ij}$ is the number of the common neighbors between user (or item) $i$ and user (or item) $j$ in the corresponding bipartite network. Similarly, we obtain another user-user network and item-item network from the two corresponding bipartite network with category, respectively, with the weight denoted by $CA_{ij}$. This process is shown in Fig.~\ref{fig:frame} B.
\item For either users or items, the two projection networks are integrated as one network, as shown in Fig.~\ref{fig:frame} C, where the link weight $W_{ij}$ between user (or item) $i$ and user (item) $j$ is defined as
\begin{eqnarray}
W_{ij} &=& CK_{ij} \times CA_{ij}.
\label{equ:category}
\end{eqnarray}
This indicates that, by comparing with the traditional one-mode projection network just based on user-item relationships, our method can naturally integrate more information about items or users.
\end{enumerate}

\subsection{Network Representation Using Node2vec}
Though we enrich the network structure according to item categories, it is difficult to capture appropriate network features using traditional network analysis methods. We thus adopt node2vec to learn continuous feature representations of nodes in a network. Since its flexible neighborhood sampling strategy, node2vec can learn rich representation in a network, and meanwhile reduce the effect of data sparsity on the recommendation algorithm. Here, we use it to automatically capture network features of the generated projection networks to transform each user (or item) into a vector.

\subsection{Clustering Users and Items by SCDNN}
After transforming the users and items into vectors, the next important step is clustering them. In this paper, we use the SCDNN method, which is based on DNNs and automatically cluster number determination algorithm, to cluster users (items) into several clusters based on the related user (item) vectors, as is described from the following two aspects.

\emph{First, we construct the DNNs similarity matrix.} Many real-world datasets are multi-scale datasets, which are of quite different distribution densities of the user (item) in different clusters. Many clustering methods are difficult to obtain good clustering results on such multi-scale datasets.

Considering the above problem, Zelnik and Perona~\cite{zelnik2005self} proposed a novel spectral clustering method, called self-tuning spectral clustering (STSC) method, where local-scale parameters are adopted. Its Gaussian similarity function is defined as:
\begin{equation}
S_{ij} = \exp\left[\frac{-d^2(i,j)}{2\sigma_i\sigma_j}\right],
\label{equ:STSC}
\end{equation}
where $\sigma_i=d(i, t)$ is the distance between the data point $i$ and its $t$-th nearest-neighbors.

\begin{figure}[!t]
\centering
\includegraphics[width=\linewidth]{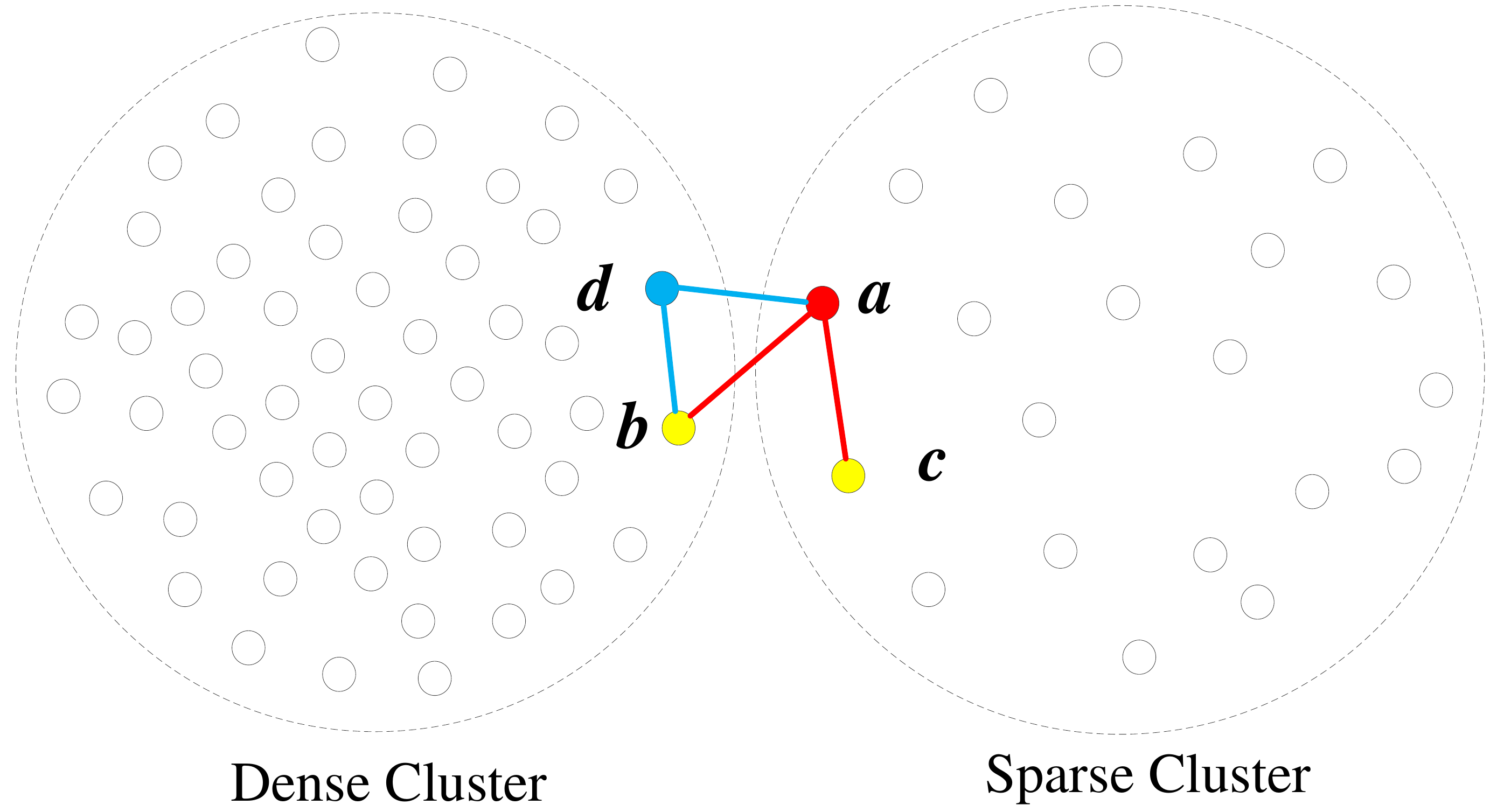}
\caption{An example of multi-scale dataset.}
\label{fig:SCDataset}
\end{figure}

In Fig.~\ref{fig:SCDataset}, based on the definition of $\sigma$, we can see $\sigma_c>\sigma_b>0$ and $\sigma_a\sigma_c>\sigma_a\sigma_b>0$. Then, based on Eq.~(\ref{equ:STSC}), we have $S_{ab}<S_{ac}$, consistent with the fact here. However, assuming $d(d, b)=d(d, a)$ in Fig.~\ref{fig:SCDataset}, based on the definition of the local-scale parameters, we can find that $\sigma_a>\sigma_b>0$ and $\sigma_a\sigma_d>\sigma_b\sigma_d>0$. Thus, according to Eq.~(\ref{equ:STSC}), the similarity between $a$ and $d$ is larger than that between $b$ and $d$, which however is inconsistent with the fact.

\begin{figure*}[!t]
\centering
\includegraphics[width=\linewidth]{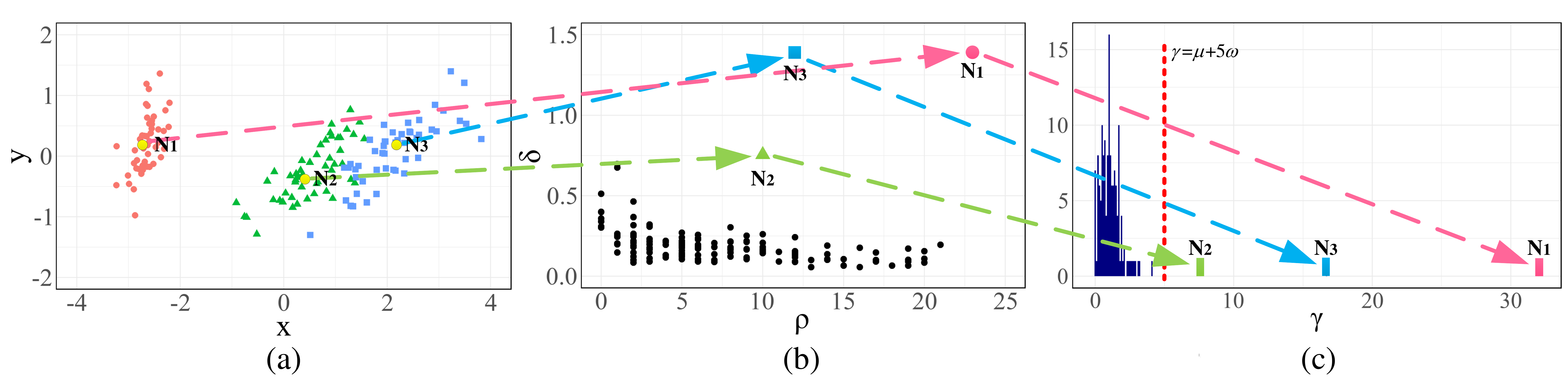}
\caption{(a) The original data distribution; (b) the density-distance $\rho-\delta$ distribution; (c) and the density distribution of $\gamma$, for a simple dataset.}
\label{fig:CNAD}
\end{figure*}

Therefore, we can find that the similarity can be indeed affected by the density difference between data points. The density of the data point $i$ can be calculated as follows:
\begin{eqnarray}
\rho_i &=& \sum_{j}{f_i(d(i,j))},
\label{equ:SCDef1_1}\\
f_i(x) &=&  \left\{
\begin{array}{rl}
1 & x \in H_i \\
0 & x \notin H_i
\end{array}\right.
\label{equ:SCDef1_2}
\end{eqnarray}
where $H_i$ is a set containing the first $p$ shortest distances between $i$ and other data points.

Considering the above analysis, we propose a novel similarity function, namely DNNs similarity function, defined by:
\begin{eqnarray}
S_{ij} &=&  \left\{
\begin{array}{ll}
\exp (\frac{-d^2(i,j)}{2\left[\max\{ \sigma_i,\sigma_j \}\right]^2}) & j\in T_i \\
0 & j \notin T_i
\end{array}\right.
\label{equ:SCDef3_1}\\
\sigma_i &=& \sum_{j \in T_i}{\frac{d(i,j)}{|T_i|}}.
\label{equ:SCDef3_2}
\end{eqnarray}
The local-scale parameter $\sigma_i$ of data point $i$ is determined by the average distance between the data point $i$ and its DNNs $j\in T_i$ defined as:
\begin{eqnarray}
T_i &=& \{ j \in N_i \vert d(i,j) < \min_{k \in J_i}(d(k,i)) \}
\label{equ:SCDef2_1}, \\
J_i &=& \{ j \in N_i \vert | \rho_i - \rho_j| > \theta \}
\label{equ:SCDef2_2},
\end{eqnarray}
where $N_i$ is the initial neighbor set~\cite{xiang2008spectral} for the data point $i$, and $\theta$ represents the density difference threshold.

In Fig.~\ref{fig:SCDataset}, we can find that $b$ and $d$ are both located in the dense cluster, while $a$ and $c$ are both located in the sparse cluster. Assuming that $|\rho_a-\rho_b|>\theta$, $|\rho_a-\rho_d|>\theta$, $|\rho_a-\rho_c|<\theta$, and $|\rho_d-\rho_b|<\theta$, based on the definition of the DNN set, we can conclude that $a \notin T_b$, $a \notin T_d$, $c \in T_a$, and $b \in T_d$. Therefore, according to Eqs.~(\ref{equ:SCDef3_1}) and (\ref{equ:SCDef3_2}), we get $S_{ac}>S_{ab}=0$ and $S_{bd}>S_{ad}=0$, which are now consistent with the fact, indicating the effectiveness of this new definition of similarity.



\emph{Second, for automatically determining the cluster number, we use the automatically determining cluster number (ADCN) method.} Based on the fast density clustering algorithm~\cite{jinyin2017novel}, in our proposed ADCN method, the cluster centers are automatically determined by constructing a normal distribution for density-distance mapping to figure out all the cluster centers.

\textbf{Definition 3}: The \emph{minimum distance} of data point $i$ is defined as the minimum distance between it and the data points of higher density, as defined by:
\begin{equation}
\delta_i = \left\{
\begin{array}{ll}
\min \{ D_h(i) \} & \rho_i\neq\max(\rho) \\
\max(\delta) & \rho_i = \max(\rho)
\end{array}\right.
\label{equ:SCDef4_1}
\end{equation}
where the set $D_h(i)$ is composed of the data points that have higher density than data point $i$.

Now, based on Eq.~(\ref{equ:SCDef1_1}) and Eq.~(\ref{equ:SCDef4_1}), we can calculate the density-distance (\(\rho-\delta\)) distribution. As an example, Fig.~\ref{fig:CNAD} (a) and (b) give the original data distribution and $\rho-\delta$ distribution, respectively, for a simple dataset, where we can see that cluster centers, represented by $N_1$, $N_2$ and $N_3$, have relatively larger $\rho$ and $\delta$ than the other data points.

Based on the density-distance distribution, we further introduce a variable $\gamma_i$ for each data point $i$, defined as
\begin{equation}
\gamma_i = \rho_i \times \delta_i,
\label{equ:gamma}
\end{equation}
to determine the cluster centers more automatically. The density distribution of $\gamma$ is shown in Fig.~\ref{fig:CNAD} (c), where we can see that it is close to a normal distribution. The cluster centers have relatively larger $\gamma$ than the other data points, which thus can be considered as the exceptional data points. Suppose, the mean value of $\gamma_i$ is $\mu$, and the variance is $\omega^2$. Based on the pauta criterion~\cite{zhang1997the}, i.e., the probability that $\gamma_i$ falls into the confidence interval $[\mu-3\omega,\mu+3\omega]$ is 99.73\%. Since the number of users or number of items in a recommender system is typically quite large, we enlarge the confidence interval to $[\mu-5\omega,\mu+5\omega]$ so that the probability that $\gamma_i$ falls into this interval is close to 99.99999999\%. In this case, we have high confidence to consider that almost all the data points are contained in the interval $[\mu-5\omega,\mu+5\omega]$, while those exceptions, i.e., $\gamma_i>\mu+5\omega$, are cluster centers. In particular, we use the following two steps to automatically determine the cluster centers: 1) Calculate the mean value $\mu$ and the variance $\omega^2$ of $\gamma$; 2) Treat the data points with $\gamma_i>\mu+5\omega$ as the cluster centers, e.g., $N_1$, $N_2$, $N_3$ in Fig.~\ref{fig:CNAD} (c).

\subsection{Two-phase Personalized Recommendation}
After clustering users and items according to their vectors and the SCDNN algorithm, we first recommend item clusters to user clusters, and then further realize the personalized recommendation of items for each user. We call it a two-phase personalized recommendation, which is described as follows.
\begin{enumerate}
\item First, we use the number of user-item relationships between each item clusters and the target user cluster to quantify the item cluster. Then, we use a basic clustering method, K-means method, to divide all item clusters into two classes, and recommend the item clusters in the class with the larger average weights to the user cluster.
\item Based on above clustering recommendation results, some traditional recommendation algorithms are adopted to recommend items in the related item clusters to the users in each user cluster, based on the related rating records.
\end{enumerate}

\section{Experiments\label{Exp}}
The proposed method is tested on multiple real-world datasets. In this paper, the node2vec~\cite{grover2016node2vec} method is implemented in Python, the SCDNN method is implemented in Matlab and the conventional recommendation methods are implemented in R. In this section, we first introduce the datasets and the recommendation algorithms for comparison. Meanwhile, we also visualize the networks established in our framework to help readers better understand our N2VSCDNNR. Finally, we give the experimental results with explanations.

\subsection{Datasets}
The real-world datasets used in the experiments are described as follows, with their basic statistics summarized in TABLE~\ref{table:dataset}.
\begin{itemize}
\item \textbf{Yelp:} The Yelp dataset includes the user reviews on Yelp from 11 cities across 4 countries. Here, two American cities, i.e., Pittsburgh and Madison, are chosen, and the reviews are utilized to define user-item interactions. There are 161 different categories of items for Pittsburgh, and 150 categories of items for Madison.
\item \textbf{Amazon:} the Amazon dataset contains the user reviews on Amazon over 26 types of goods. Here, Musical Instruments dataset is chosen and the reviews are utilized to define user-item interaction. There are total 912 categories of items.
\item \textbf{MovieLens:} MovieLens dataset contains the user ratings over movies on MovieLens. the ratings are utilized to define user-item interaction. Note that it only contains the users with more than 20 ratings and demographic information. There are total 50 categories of movies.
\end{itemize}

\begin{table}[ht]
\centering
\caption{Statistics of the four datasets.}
\begin{tabular}{@{}ccccc@{}}
\hline \hline
     \textbf{Dataset}   & \textbf{\#User}   & \textbf{\#Item} & \textbf{\#Link}   & \textbf{\#Category} \\ \midrule
  Yelp (Pittsburgh) & 466 & 1,672    & 10,373 & 161  \\
  Yelp (Madison) & 332 & 1172    & 5,597 & 150  \\
  Amazon & 6831 & 32054    & 71,661 & 912  \\
  MovieLens & 943 & 1,682 & 100,000 & 50  \\ \hline \hline
\end{tabular}
\label{table:dataset}
\end{table}

\begin{figure*}[!t]
\centering
\includegraphics[width=\linewidth]{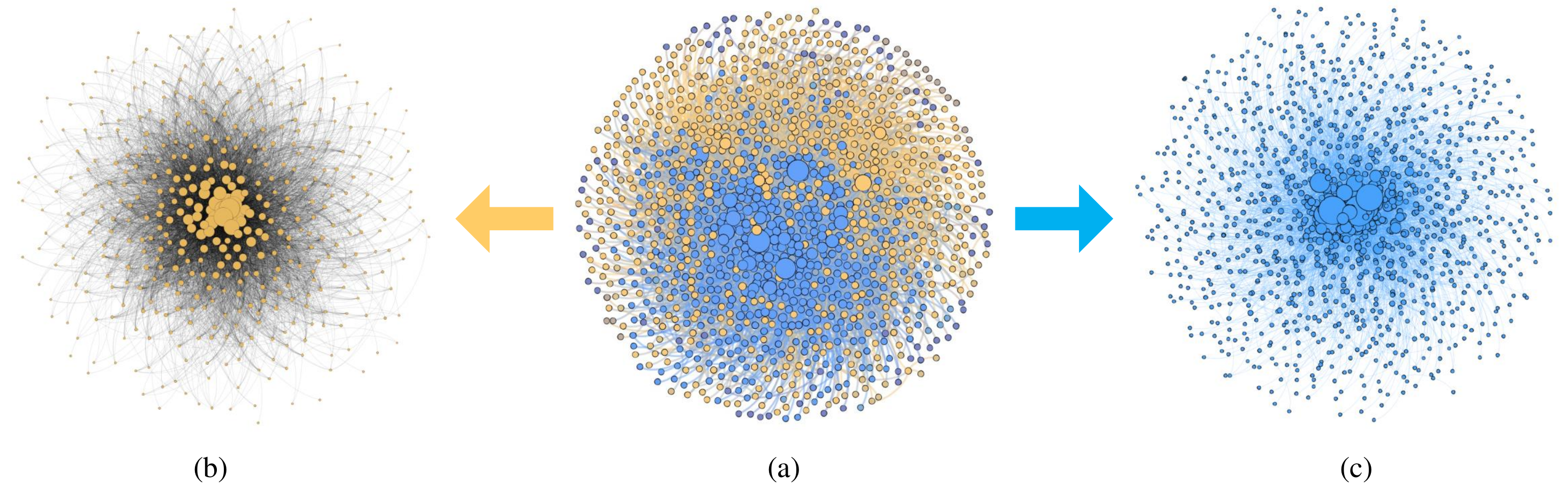}
\caption{Based on the one-mode projection of (a) the original user-item bipartite network, we obtain (b) the user-user projection network and (c) the item-item projection network.}
\label{fig:ProjectionNet}
\end{figure*}

\subsection{Models and Algorithms for Comparison\label{Comparison}}
In order to evaluate the proposed framework, we compare our method with the following two models.
\begin{itemize}
\item \textbf{Original:} We directly use the traditional recommendation algorithms.
\item \textbf{N2VSCDNN:} This method was proposed in ~\cite{chen2017improved}. As the previous version of N2VSCDNNR, it is purely based on user-item interactions without any category information.
\end{itemize}

In the two-phase personalized recommendation, we use the following four popular recommendation algorithms.
\begin{itemize}
\item \textbf{UBCF~\cite{zhao2010user}:} User-based collaborative filtering (UBCF) first finds out the users similar to the target user and then recommend items based on these similar users.
\item \textbf{IBCF~\cite{sarwar2001item}:} Item-based collaborative filtering (IBCF) is a kind of collaborative filtering algorithms based on the similarity between items calculated by using the ratings of these items.
\item \textbf{NMF~\cite{ding2010convex}:} Non-negative matrix factorization (NMF) is a group of algorithms based on multivariate analysis. It seeks to approximate the input matrix with the multiplication of the $d$-dimensional low-rank representations. In this paper, we set the dimension $d=40$.
\item \textbf{Popular:} It divides the subset of users and items according to certain rules or attributes, and then recommends the items with the highest popularity to the users.
\end{itemize}

To validate the effectiveness of our N2VSCDNNR, we choose two advanced embedding-based, as well as a side-information-based~\cite{gopalan2014content,porteous2010bayesian,do2018coupled,hu2016non,gao2018bine}, recommendation algorithms for comparison, which are briefly described as follows.
\begin{itemize}
\item \textbf{Metapath2vec++~\cite{dong2017metapath2vec}:} This is the state-of-the-art method for embedding heterogeneous networks. The meta-path scheme chosen in our experiments is item-user-item.
\item \textbf{BiNE~\cite{gao2018bine}:} As a novel network representation method, it was proposed to learn the representations for bipartite networks. It jointly models both the explicit relations and high-order implicit relations in learning the representation for nodes.
\item \textbf{CoFactor~\cite{Liang2016Factorization}:} This is a co-factorization model inspired by word2vec~\cite{levy2014neural,mikolov2013distributed}, which jointly decomposes the user-item interaction matrix and the item-item co-occurrence matrix with shared item latent factors.
\end{itemize}

For each network representation learning method, we use the released implementations of the authors for our experiments, and adopt the inner product kernel $u_i^Tv_j$ to estimate the preference of user $u_i$ on item $v_j$, and evaluate performance on the top-ranked results.

In this paper, we use 5-fold cross-validation to evaluate the performances of the methods, based on the four basic measurements in Top-N recommendation, including precision, recall, Hit Rank (HR) and Average-Reciprocal Hit Rank (ARHR).

As indicated in \cite{rodriguez2014clustering}, when the parameter $p$ in the ADCN method is $1\%\sim2\%$ of the dataset, we can obtain good clustering result. Therefore, we determine the clusters number when $p=2\%$. Moreover, to better reflect the network structure, we set the feature representation dimension $d=100$, and set the in-out and return parameters both to be 1 in the node2vec algorithm.

\subsection{Network Visualization}
In order to provide a more intuitive view of our method, we visualize the networks generated in the process of analyzing the Yelp (Madison) dataset as an example.

First, the user-item bipartite network is shown in Fig.~\ref{fig:ProjectionNet} (a), where users and items are denoted by yellow and blue nodes, respectively. The node size is proportional to its degree in the network. Then, the bipartite network is compressed by one-mode projection to get the corresponding user projection network and item projection network, as shown in Fig.~\ref{fig:ProjectionNet} (b) and (c), respectively.

\begin{figure}[!t]%
\centering
\includegraphics[width=\linewidth]{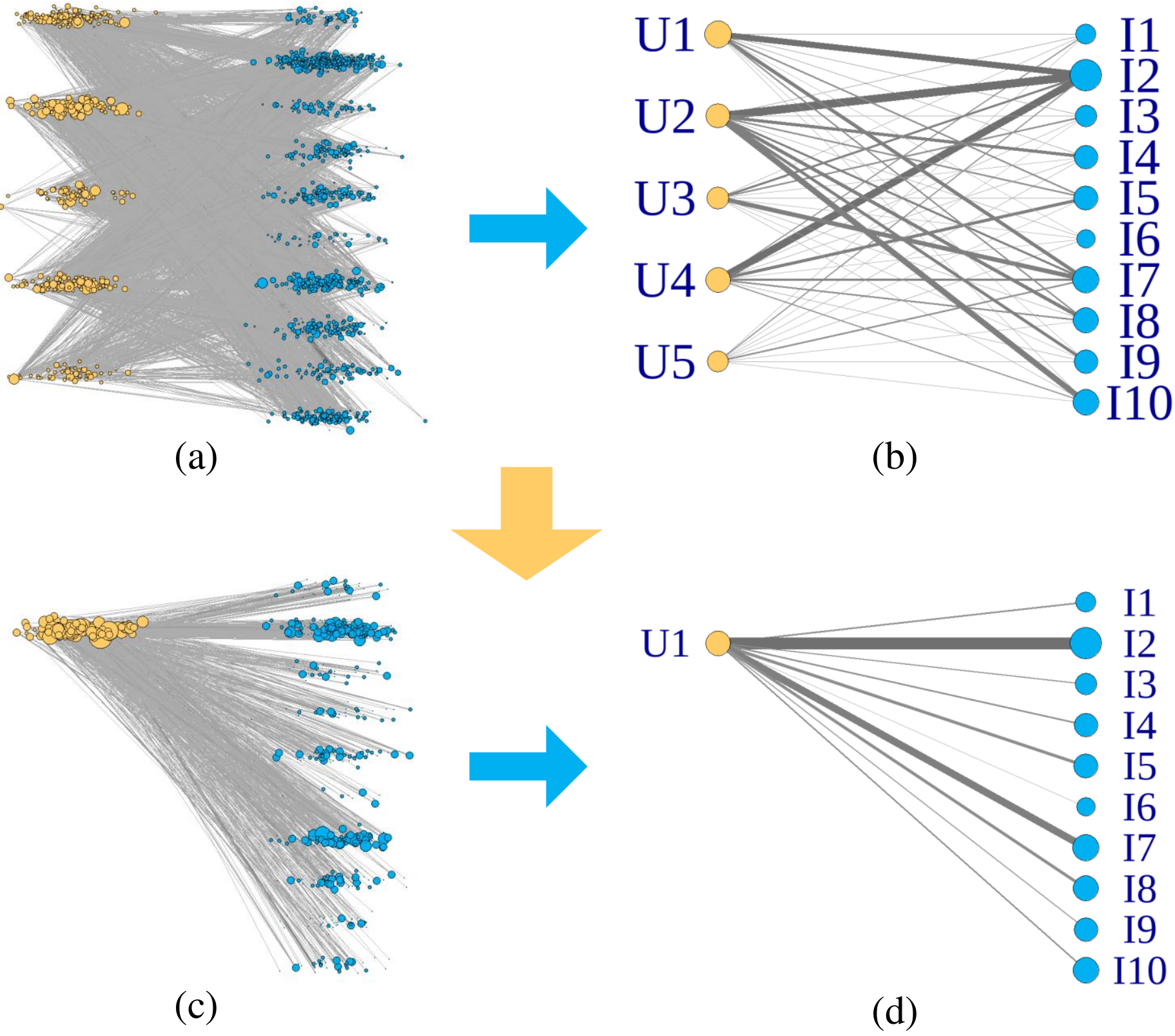}
\caption{(a) The original user-item bipartite network with the user and item nodes clustered based on the corresponding node vectors. (b) The weighted bipartite network between user clusters and item clusters. (c) The user-item bipartite subnetwork by only considering the user nodes in the cluster U1. (d)  The weighted bipartite network between the user cluster U1 and all the item clusters.}%
\label{fig:ClusterNet}%
\end{figure}

Next, the node2vec algorithm is applied to transform the nodes in user (item) projection category network into user (item) vectors. And the SCDNN algorithm is used to cluster users (items), as shown in Fig.~\ref{fig:ClusterNet} (a), based on which a weighted bipartite network is generated between user (item) clusters, as shown in Fig.~\ref{fig:ClusterNet} (b), where user (item) clusters are denoted by yellow (blue) nodes. The node size is proportional to the number of the users (items) in the cluster. Each link is weighted by the total number of the relationships between the users (items) in the corresponding user (item) clusters.

Take user cluster U1 for example, the relationships between it and all the item clusters are shown in Fig.~\ref{fig:ClusterNet} (c)-(d), where we can see that U1 has relatively stronger relationships with the item clusters I2 and I7 than the others. K-means algorithm then is used to divide all item clusters into two classes and recommend those in the class with larger average weight to the target user cluster U1, i.e., here we recommend I2 and I7 to U1. Finally, different recommendation algorithms are used to recommend items in I2 and I7 to each user in U1, according to their ratings.

\subsection{Results}
\begin{figure*}[!t]
\centering
\subfigure[Yelp (Pittsburgh)]{
\includegraphics[width=.49\linewidth]{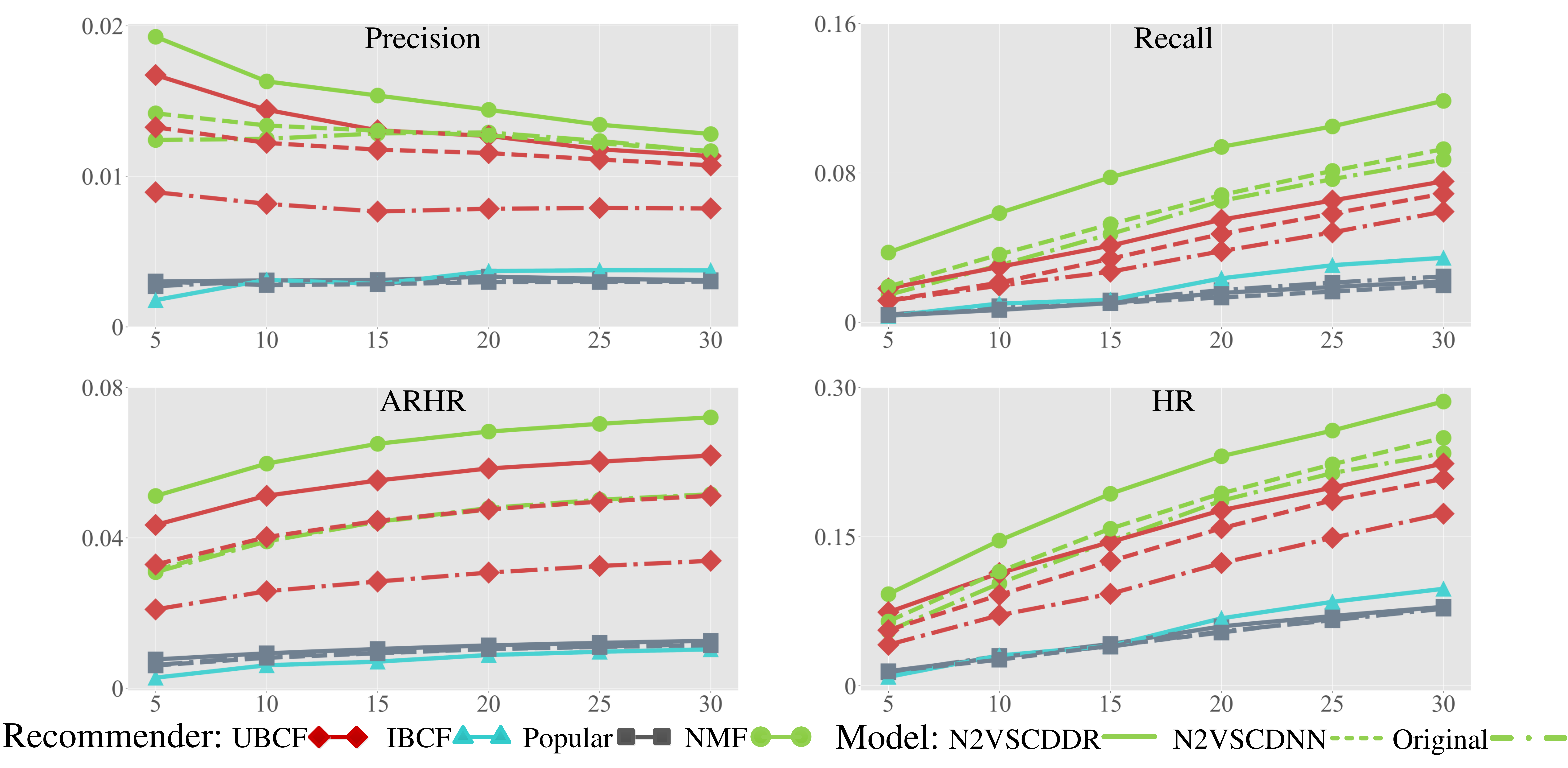}}
\subfigure[Yelp (Madison)]{
\includegraphics[width=.49\linewidth]{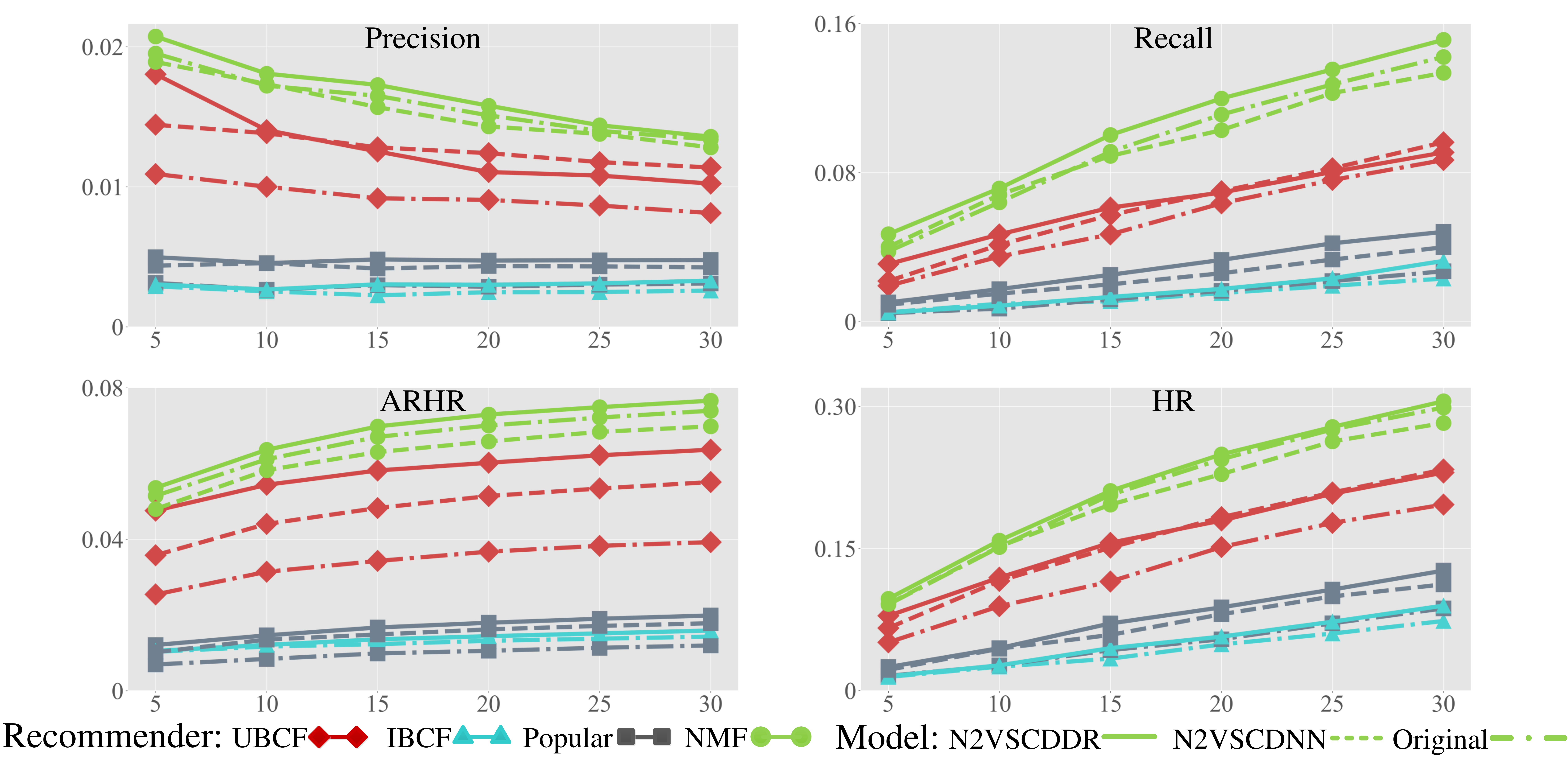}}
\subfigure[Amazon]{
\includegraphics[width=.49\linewidth]{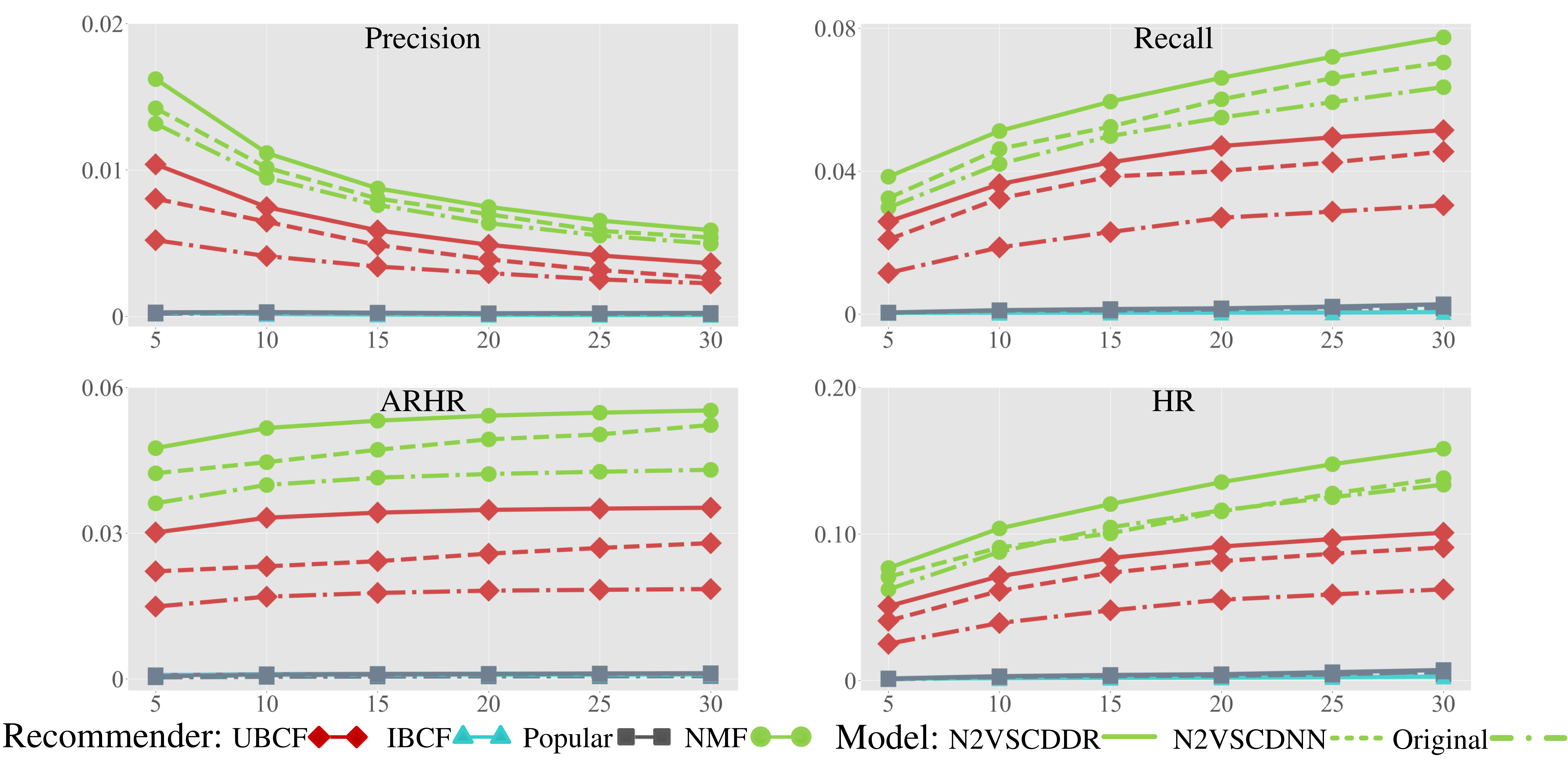}}
\subfigure[MovieLens]{
\includegraphics[width=.49\linewidth]{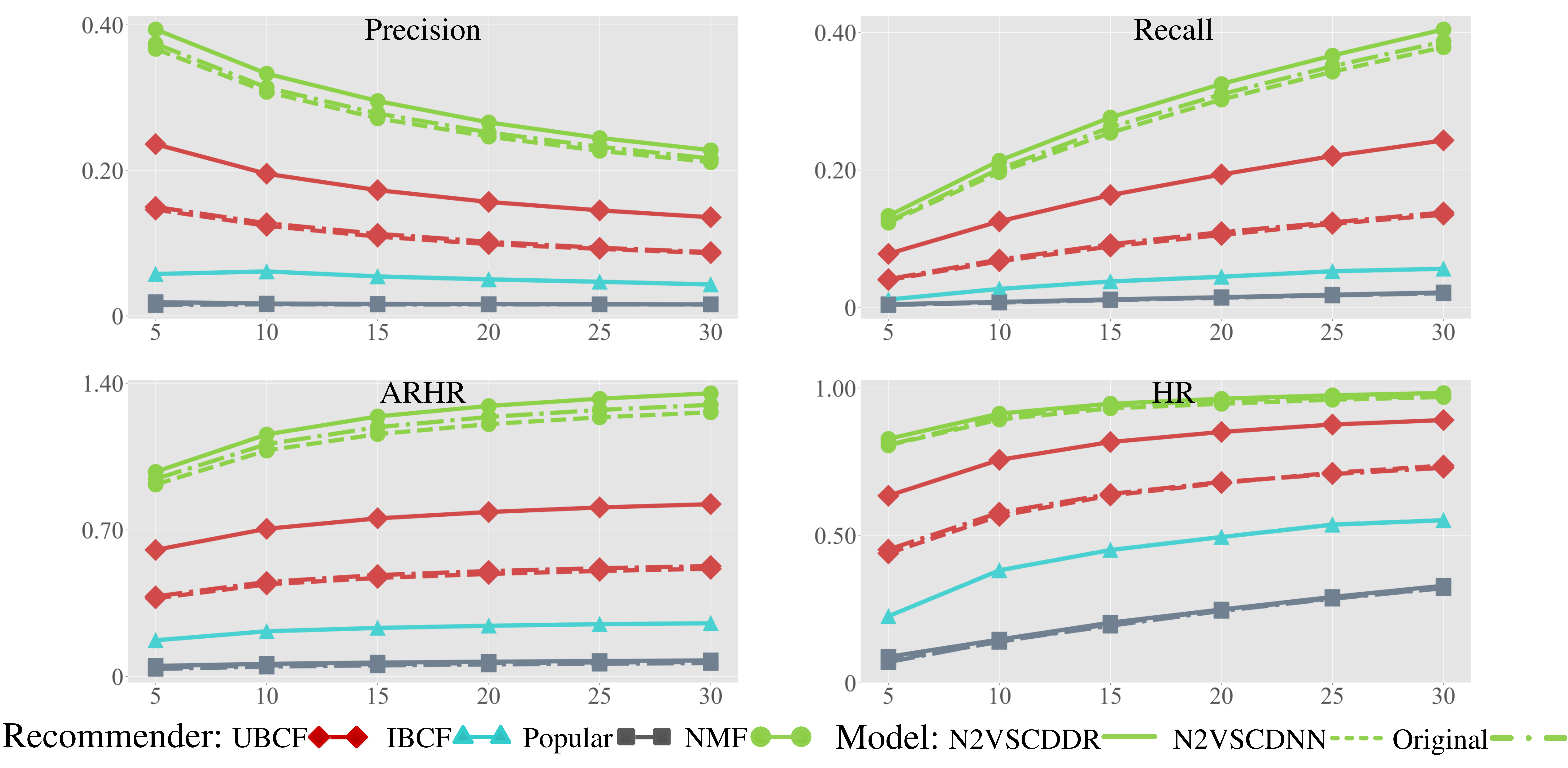}}
\caption{The recommendation results of NMF-based recommender systems with proposed models and three baselines on the four datasets: (a) Yelp (Pittsburgh), (b) Yelp (Madison), (c) Amazon, and (d) MovieLens.}
\label{original}
\end{figure*}

In this section, at first, we compare the results obtained by the three models, including Original, N2VSCDNN and N2VSCDNNR, based on the four basic recommendation algorithms, including NMF, UBCF, IBCF and Popular, as shown in Fig.~\ref{original}. We can see that, in general, our method N2VSCDNNR behaves better than N2VSCDNN, both of which behaves better than Original, in almost all the cases, by adopting any basic recommendation algorithm and using any performance measurement. Such superiorities are quite significant for the first three datasets, i.e., Yelp (Pittsburgh), Yelp (Madison), and Amazon, by adopting the NMF and UBCF recommendation algorithms. However, they are significant for the MovieLens only when the UBCF is adopted. Meanwhile, when comparing the four basic recommendation algorithms, the NMF behaves the best, the UBCF follows, while the IBCF and the Popular behave the worst, by using any model on any dataset.

\begin{figure*}[!t]
\centering
\subfigure[Yelp (Pittsburgh)]{
\includegraphics[width=.49\linewidth]{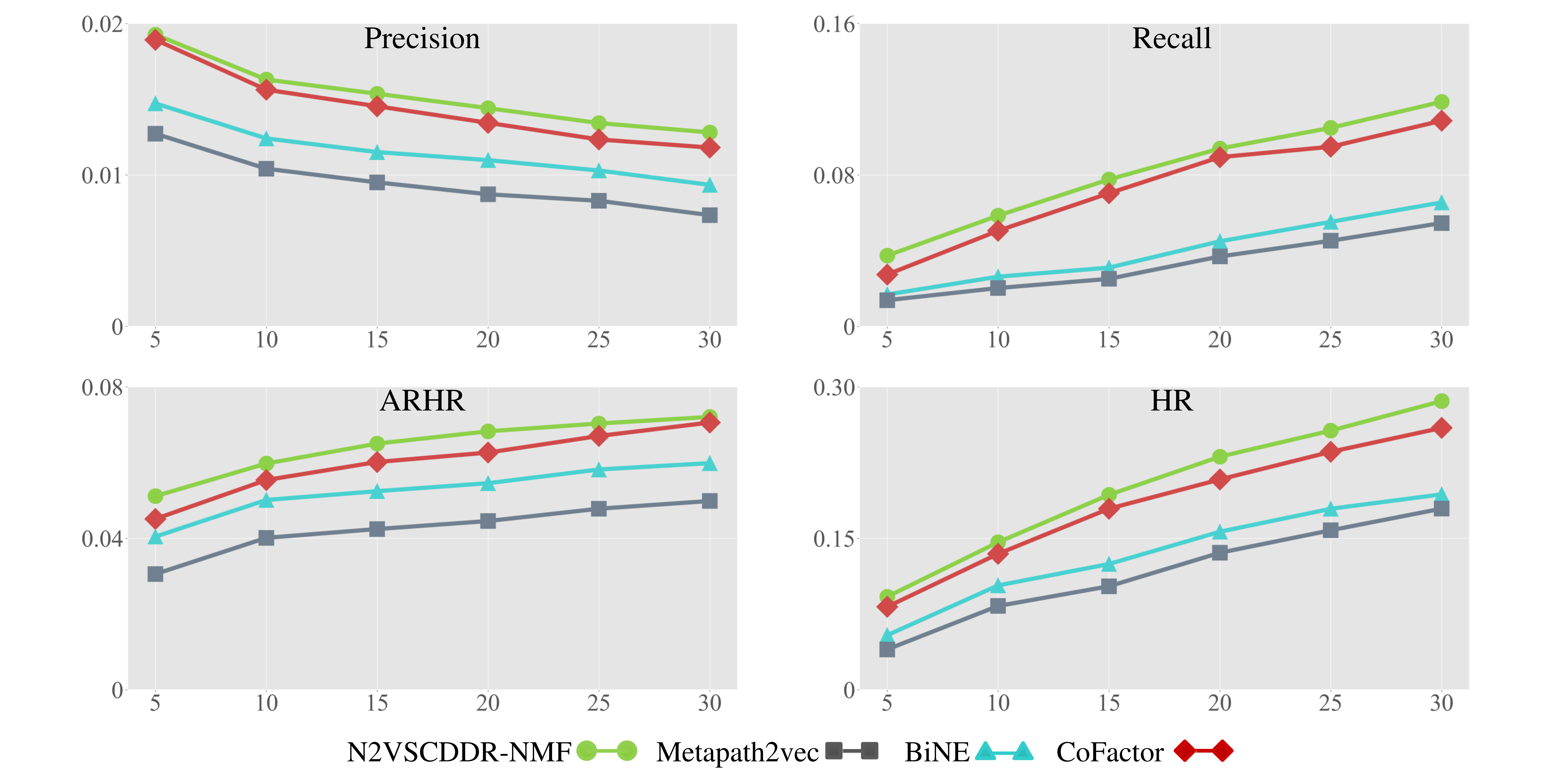}}
\subfigure[Yelp (Madison)]{
\includegraphics[width=.49\linewidth]{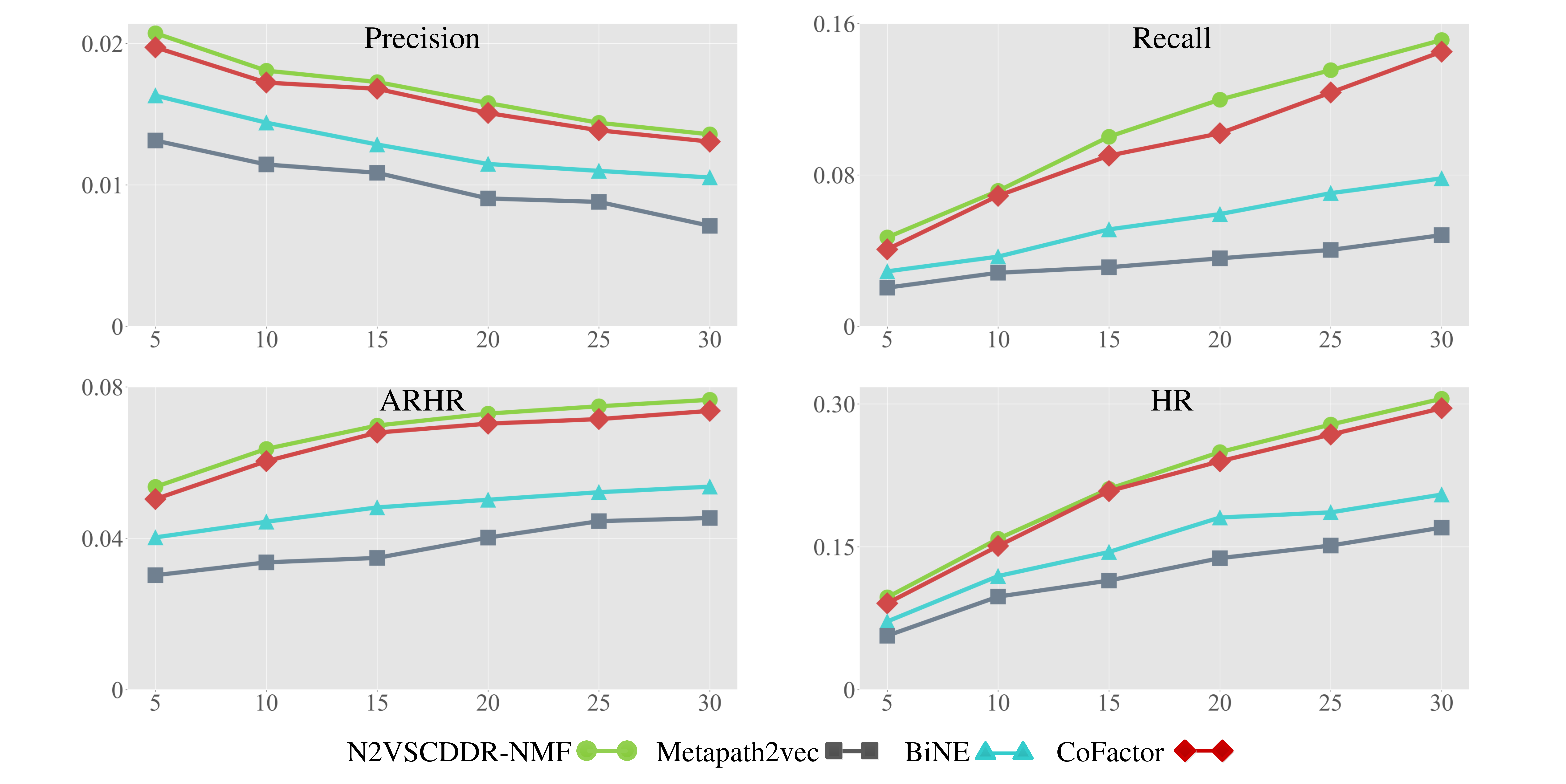}}
\subfigure[Amazon]{
\includegraphics[width=.49\linewidth]{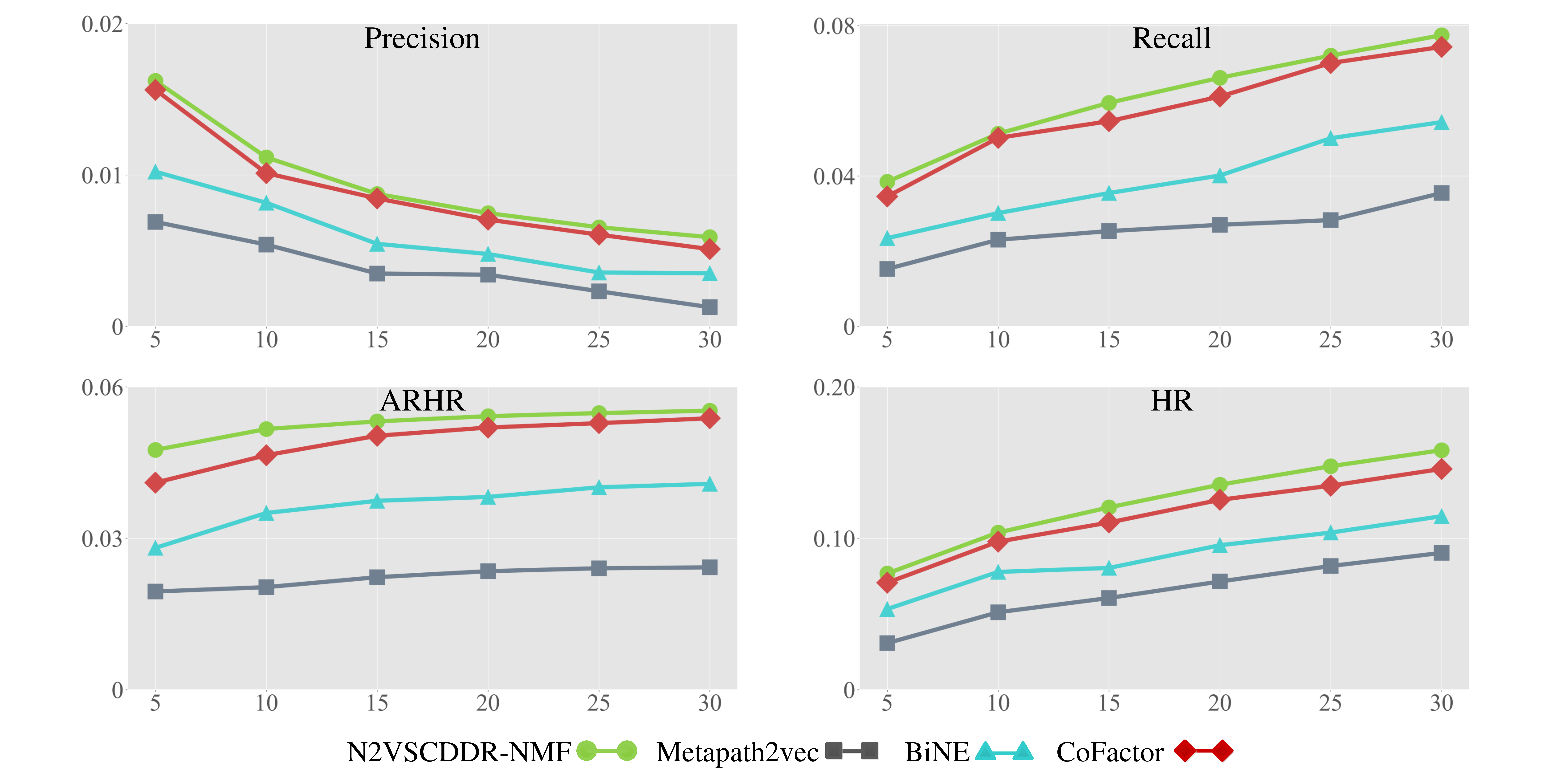}}
\subfigure[MovieLens]{
\includegraphics[width=.49\linewidth]{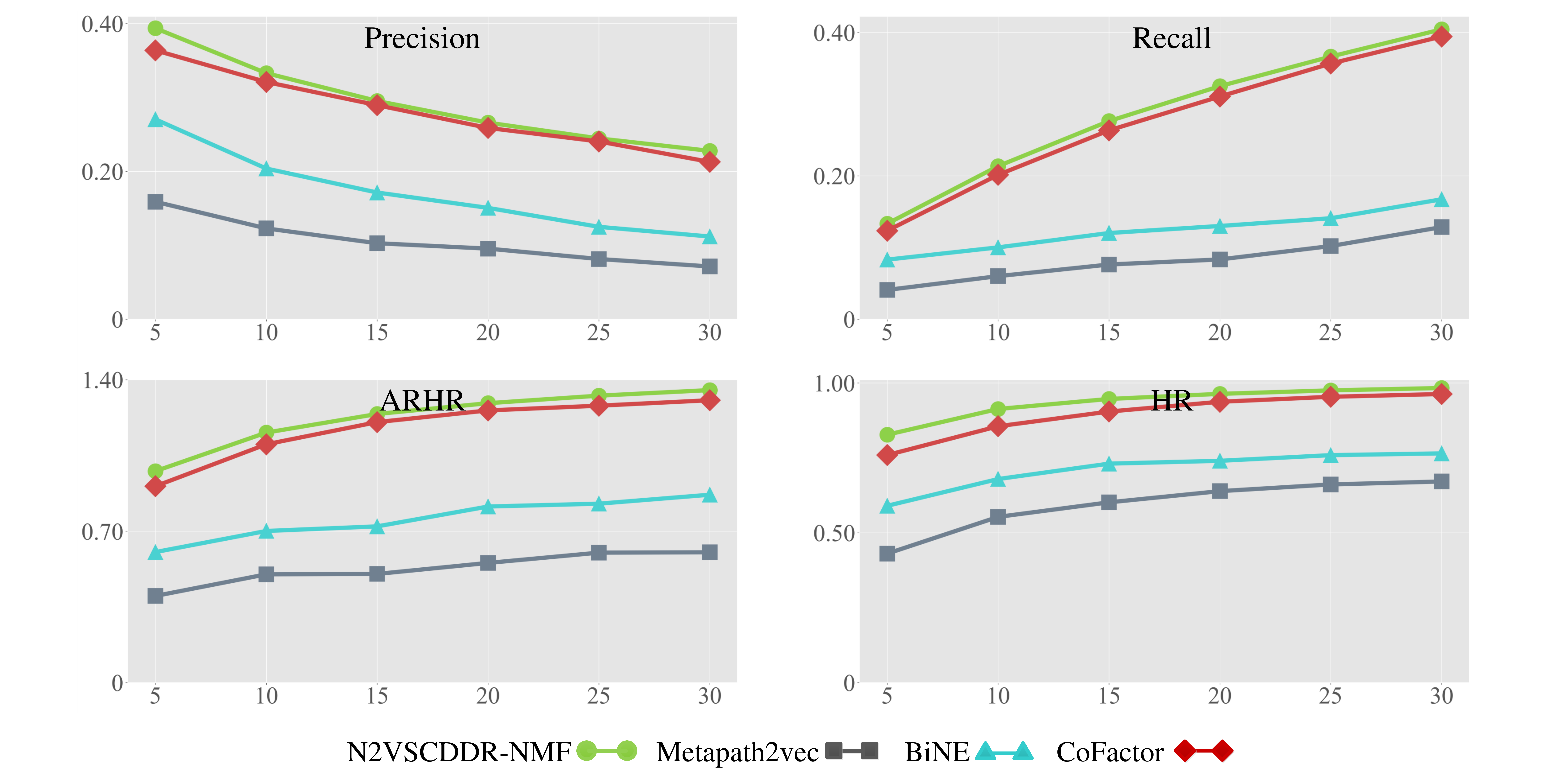}}
\caption{The recommendation results of various baselines on the four datasets: (a) Yelp (Pittsburgh), (b) Yelp (Madison), (c) Amazon, and (d) MovieLens.}
\label{baseline}
\end{figure*}

In the following, we thus mainly focus on the NMF and UBCF recommendation algorithms, and try to reveal the relatively performance improvements by using N2VSCDNN and N2VSCDNNR, respectively, comparing with the Original model. The results are presented in TABLE~\ref{table:yelp_3}-\ref{table:movie_table}. Overall, larger relatively improvements are obtained by using our N2VSCDNNR, in most cases, for each performance metric. Consistent with the intuitive pictures of Fig~\ref{original}, for the three datasets including Yelp (Pittsburgh), Yelp (Madison), and Amazon, such improvements are relatively significant for both the NMF and UBCF recommendation algorithms, while for the MovieLens, such improvements are remarkable only when our N2VSCDNNR model and the UBCF recommendation algorithm are adopted together.

\begin{table}[!t]
\centering
\caption{The average relative improvements of performances (\%) introduced by N2VSCDNN and N2VSCDNNR on Yelp (Pittsburgh), by adopting the NMF and the UBCF recommendation algorithms.}
\begin{tabular}{cccccc}
\hline \hline
\textbf{Algorithm}  & \textbf{Model} & \textbf{ARHR} & \textbf{HR} & \textbf{Precision} & \textbf{Recall} \\ \hline
\multirow{2}{*}{NMF}  & N2VSCDNNR  &\textbf{46.53}&\textbf{28.48}&\textbf{22.72}&\textbf{52.89}\\
& N2VSCDNN &0.00&6.99&3.41&8.94                 \\  \hline
\multirow{2}{*}{UBCF} & N2VSCDNNR &\textbf{93.84}&\textbf{71.40}&\textbf{77.92}&\textbf{81.45}         \\
& N2VSCDNN &43.73&50.59&42.00&57.75            \\ \hline \hline
\end{tabular}
\label{table:yelp_3}
\end{table}

\begin{table}[!t]
\centering
\caption{The average relative improvements of performances (\%) introduced by N2VSCDNN and N2VSCDNNR on Yelp (Madison), by adopting the NMF and the UBCF recommendation algorithms.}
\begin{tabular}{cccccc}
\hline \hline
\textbf{Algorithm}  & \textbf{Model} & \textbf{ARHR} & \textbf{HR} & \textbf{Precision} & \textbf{Recall} \\ \hline
\multirow{2}{*}{NMF}  & N2VSCDNNR &\textbf{3.96}&\textbf{2.49}&\textbf{4.33}&\textbf{8.99}\\
& N2VSCDNN&-5.69&-4.25&-2.99&-3.03                 \\  \hline
  \multirow{2}{*}{UBCF} & N2VSCDNNR&\textbf{68.58}&\textbf{24.57}&\textbf{40.47}&\textbf{15.91}   \\
& N2VSCDNN&40.26&22.84&36.94&12.68            \\ \hline \hline
\end{tabular}
\label{table:yelp_5}
\end{table}

\begin{table}[!t]
\centering
\caption{The average relative improvements of performances  (\%) introduced by N2VSCDNN and N2VSCDNNR on Amazon, by adopting the NMF and the UBCF recommendation algorithms.}
\begin{tabular}{cccccc}
\hline \hline
\textbf{Algorithm}  & \textbf{Model} & \textbf{ARHR} & \textbf{HR} & \textbf{Precision} & \textbf{Recall} \\ \hline
\multirow{2}{*}{NMF}  & N2VSCDNNR &\textbf{29.00}&\textbf{17.92}&\textbf{18.94}&\textbf{21.77}\\
& N2VSCDNN&16.57&2.19&7.48&9.42                 \\  \hline
\multirow{2}{*}{UBCF} & N2VSCDNNR&\textbf{93.84}&\textbf{71.40}&\textbf{77.92}&\textbf{81.45}         \\
& N2VSCDNN&43.73&50.59&42.00&57.75            \\ \hline \hline
\end{tabular}
\label{table:amazon_table}
\end{table}

\begin{table}[!t]
\centering
\caption{The average relative improvements of performances (\%) introduced by N2VSCDNN and N2VSCDNNR on MovieLens, by adopting the NMF and the UBCF recommendation algorithms.}
\begin{tabular}{cccccc}
\hline \hline
\textbf{Algorithm}  & \textbf{Model} & \textbf{ARHR} & \textbf{HR} & \textbf{Precision} & \textbf{Recall} \\ \hline
\multirow{2}{*}{NMF}  & N2VSCDNNR &\textbf{4.16}&\textbf{1.22}&\textbf{5.60}&\textbf{5.07}\\
& N2VSCDNN&-2.77&-0.65&-2.17&-2.36                 \\  \hline
\multirow{2}{*}{UBCF} & N2VSCDNNR&\textbf{56.32}&\textbf{27.40}&\textbf{54.74}&\textbf{78.25}   \\
& N2VSCDNN&-2.56&-0.55&-2.51&-2.88            \\ \hline \hline
\end{tabular}
\label{table:movie_table}
\end{table}

Quite impressively, when we adopt our N2VSCDNNR model based on the UBCF recommendation algorithm, we can get huge improvements based on any performance metric, e.g., they are even close to 100\%  for the Yelp (Pittsburgh) and Amazon datasets. This indicates that our model is especially useful to enhance the efficiency of user-based collaborative filtering method. Although the improvements introduced by our N2VSCDNNR model seems relatively small when the NMF is adopted, but they are still larger than those introduced by the N2VSCDNN model. This is mainly because the NMF recommendation algorithm itself behaves quite well even based on the Original model, and the potential for further improvement thus is relatively low.

As we can see, the recommendation results obtained by N2VSCDNNR based on NMF are better than those based on the other basic recommendation algorithms, we thus further compare the results obtained by N2VSCDNNR-NMF with the results obtained by the three advanced embedding and side information based recommendation algorithms, including Metapath2vec++, BiNE and CoFactor. The results are presented in Fig.~\ref{baseline} and TABLE~\ref{all_improvement_baselines}, where we can see that the N2VSCDNNR-NMF outperforms all of the baseline methods in all the cases, while by comparison, Metapath2vec++ performs the worst in most cases. This may be because Metapath2vec++ just treats the explicit and implicit relations equally while ignores their weights which are useful to distinguish the importance of various relations.

\begin{table}[!t]
\centering
\caption{The average performances of N2VSCDNNR-NMF and the three baselines on the four datasets.}
\resizebox{\linewidth}{!}{
\begin{tabular}{cccccc}
\hline \hline
\textbf{Algorithm}  & \textbf{Model} & \textbf{ARHR} & \textbf{HR} & \textbf{Precision} & \textbf{Recall} \\ \hline
\multirow{4}{*}{Yelp (Pittsburgh)}  & N2VSCDNNR  &\textbf{0.0644}&\textbf{0.2008}&\textbf{0.0153}&\textbf{0.0818}\\
& Metapath2vec++ &0.0426&0.1164&0.0095&0.0326                \\
& BiNE           &0.0526&0.1352&0.0115&0.0399                \\
& CoFactor       &0.0602&0.1832&0.0144&0.0735                \\
\hline
\multirow{4}{*}{Yelp (Madison)}  & N2VSCDNNR &\textbf{0.0686}&\textbf{0.2166}&\textbf{0.0166}&\textbf{0.1042}\\
& Metapath2vec++ &0.0381&0.1214&0.0101&0.0341                 \\
& BiNE           &0.0481&0.1512&0.0128&0.0542                 \\
& CoFactor       &0.0657&0.2088&0.0160&0.0951                 \\
\hline
\multirow{4}{*}{Amazon}  & N2VSCDNNR &\textbf{0.0528}&\textbf{0.1237}&\textbf{0.0093}&\textbf{0.0608}\\
& Metapath2vec++&0.0223&0.0643&0.0038&0.0257          \\
& BiNE          &0.0366&0.0875&0.0059&0.0389          \\
& CoFactor      &0.0494&0.1141&0.0087&0.0575          \\
\hline
\multirow{4}{*}{MovieLens}  & N2VSCDNNR &\textbf{1.2241}&\textbf{0.9350}&\textbf{0.2933}&\textbf{0.2865}\\
& Metapath2vec++&1.1761&0.8957&0.2810&0.2750              \\
& BiNE          &0.7559&0.7108&0.1721&0.1236              \\
& CoFactor      &1.1761&0.8957&0.2810&0.2750              \\
\hline \hline
\end{tabular}}
\label{all_improvement_baselines}
\end{table}

\subsection{Time Complexity}
Now, let's analyze the time complexities of N2VSCDNNR and the baselines. We regard the procedures before personalized recommendation as pre-training, and only focus on the time complexity of online personalized recommendation. In particular, since N2VSCDNNR behaves the best when it is based on NMF, here, we just give the complexity of N2VSCDNNR-NMF for simplicity. Suppose the number of users is $n$, the number of items is $m$, the average number of items in the selected item clusters is $A_{vg} (A_{vg}<m)$, and the number of user consumption records is $C$. In BiNE and Metapath2vec++, the window size is $w$, and the iterations number is $Iter$. The time complexities of all the considered recommender systems are presented in TABLE~\ref{Complexity}, where we can find that our N2VSCDNNR-NMF has much lower time complexity than other recommender systems when making online recommendations, especially when we divide items into more clusters while only recommend a small number of them to the target user cluster. This indicates that N2VSCDNNR could be more suitable to be applied in large-scale systems.

\begin{table}[!h]
\centering
\caption{The time complexity of the considered recommender systems.}
\label{Complexity}
\begin{tabular}{cc}
\hline\hline
\textbf{Algorithm}     & \textbf{Time complexity} \\ \hline
N2VSCDNNR-NMF &   $O(n\cdot A_{vg})$           \\
Metapath2vec++&   $O(2(n+m)\cdot w\cdot Iter+n\cdot m)$              \\
BiNE          &   $O(2(n+m)\cdot w\cdot Iter+n\cdot m)$              \\
CoFactor      &   $O(n\cdot m+C)$              \\
\hline\hline
\end{tabular}
\end{table}

\section{Conclusion\label{Conclusion}}
In this paper, we enrich the network structure based on item categories. Then, we establish one-mode user and item projection networks, and further use node2vec technology to transform each user (or item) node to a user (or item) vector. After that, we cluster users (or items) based on these vectors, according to which we establish a bipartite cluster network using an improved spectral clustering algorithm SCDNN. Based on this bipartite cluster network, for each user cluster, we keep the item clusters with the most frequent relationships with the user cluster. Finally, we use four different recommendation algorithms to recommend the items in these item clusters to each user in the user cluster. By comparing with several advanced embedding and side information based recommendation algorithms, the experiments on four real-world datasets validate the outstanding performance of our framework, in terms of both higher precision and recall. Moreover, we also analyze the time complexity of these recommendation algorithms, and find that our N2VSCDNNR has relatively lower time complexity than the others in online  recommendation, indicating its potential to be widely applied in large-scale systems.

In the future, we are interested in utilizing more network representation methods, besides the node2vec algorithm, in recommender systems, and also try to find the optimal parameters using some optimization algorithms to obtain more comprehensive results.

\bibliographystyle{IEEEtran}
\bibliography{ref1802}

\end{document}